\documentclass[sigconf]{aamas} 

\usepackage{balance} 

\setcopyright{ifaamas}
\acmConference[AAMAS '23]{Proc.\@ of the 22nd International Conference
on Autonomous Agents and Multiagent Systems (AAMAS 2023)}{May 29 -- June 2, 2023}
{London, United Kingdom}{A.~Ricci, W.~Yeoh, N.~Agmon, B.~An (eds.)}
\copyrightyear{2023}
\acmYear{2023}
\acmDOI{}
\acmPrice{}
\acmISBN{}

\acmSubmissionID{???}

\usepackage{graphicx}
\usepackage{amsmath}
\usepackage{amsthm}
\usepackage{booktabs}
\usepackage{algorithm}
\usepackage{algorithmic}

\newtheorem{theorem}{Theorem}

\newtheorem{proposition}{Proposition}
\newtheorem{assumption}{Assumption}

\usepackage[toc,page]{appendix}
\usepackage{hyperref}
\usepackage{cleveref}

\usepackage{amsmath,amsthm,bbm}
\DeclareMathOperator*{\argmin}{argmin}
\DeclareMathOperator*{\argmax}{argmax}

\newcommand{\br}{\mathit{br}}

\theoremstyle{definition}
\newtheorem*{definition*}{Definition}

\usepackage{booktabs}
\usepackage{subfig}

\providecommand{\abs}[1]{\lvert#1\rvert}

\title{Empirical Game-Theoretic Analysis for Mean Field Games}

\author{Yongzhao Wang}
\affiliation{
  \institution{University of Michigan}
  \city{Ann Arbor}
  \country{USA}}
\email{wangyzh@umich.edu}

\author{Michael P. Wellman}
\affiliation{
  \institution{University of Michigan}
  \city{Ann Arbor}
  \country{USA}}
\email{wellman@umich.edu}

\begin{abstract}
We present a simulation-based approach for solution of mean field games (MFGs), using the framework of \textit{empirical game-theoretical analysis} (EGTA)\@.
Our primary method employs a version of the double oracle, iteratively adding strategies based on best response to the equilibrium of the \textit{empirical MFG} among strategies considered so far.
We present \textit{Fictitious Play} (FP) and \textit{Replicator Dynamics} as two subroutines for computing the empirical game equilibrium.
Each subroutine is implemented with a query-based method rather than maintaining an explicit payoff matrix as in typical EGTA methods due to a representation issue we highlight for MFGs.
We test the performance of our primary method in various games and show that it outperforms directly applying FP to MFGs with either subroutine.
By introducing game model learning and regularization, we significantly improve the sample efficiency of the primary method without sacrificing the overall learning performance.
Theoretically, we prove that a Nash equilibrium (NE) exists in the empirical MFG and show the convergence of iterative EGTA to NE of the full MFG.
\end{abstract}

\keywords{Mean Field Games; Empirical Game-Theoretic Analysis; Game Model Learning}

\begin{document}

\pagestyle{fancy}
\fancyhead{}

\maketitle

\section{Introduction}

\textit{Mean field games} (MFGs) \citep{lasry2007mean,huang2006large} model strategic interactions among a conceptually infinite number of players. 
Aggregate player behavior is summarized by a distribution over states of the population, which constitutes the \textit{mean field}. 
The MFG framework reduces analysis to the characterization of the optimal behavior of a single representative player in its interactions with the full population, as represented by the mean field.
By considering the limit case of populations with mean-field structure, this framework can support game-theoretic analysis that would be intractable for a standard corresponding model of a game among a large but finite number of players.
One common interpretation of MFG models is as an approximation of an actually finite many-player game.

The complexity of analytic solution of MFGs grows with the state space, necessitating numerical approaches for all but the simplest problems. 
Several such algorithms for MFGs have recently been proposed \citep{perrin2020fictitious,perolat2021scaling,perrin2021mean}.
As befits the game nature of MFGs, some of these employ standard online game-solving algorithms, such as fictitious play (FP) \citep{robinson1951iterative,brown1951iterative}.

The methodology of \textit{empirical game-theoretic analysis} (EGTA) \citep{TuylsPLHELSG20,Wellman06} represents a general approach for building and reasoning about game models based on simulation.
EGTA works by estimating game models over restricted strategy sets, typically identified through a heuristic process as being strategically salient for analysis.
The estimated game models are called \textit{empirical games}.
An iterative form of EGTA is presented by \citet{Lanctot17} as the \textit{Policy Space Response Oracle} (PSRO) algorithm, where at each iteration, new strategies are generated through reinforcement learning (RL)\@.
There is a considerable prior literature that demonstrates the effectiveness of PSRO and variants in a range of applications \citep{wang19sywsjf,Wright19,czarnecki2020real}.

We develop an iterative framework for numerically solving MFGs with EGTA.
Under the framework, we first introduce a primary method to demonstrate the basic implementation of the framework, and then we apply game model learning (GML) and regularization to the primary method, which significantly improve the sample efficiency of the primary method without sacrificing the overall learning performance.
Our primary method extends the double oracle method \citep{mcmahan2003planning} to MFGs, iteratively adding strategies based on best response to the equilibrium of the \textit{empirical MFG} among strategies considered so far.
We propose FP and replicator dynamics (RD) as two subroutines for computing a Nash equilibrium (NE) of the empirical MFG and demonstrate that both subroutines are effective for the empirical game analysis.
Each subroutine is implemented with a query-based method, in which we query the utilities of different strategies through simulations as needed rather than maintaining an explicit payoff matrix as in typical EGTA methods.
This tweak is caused by the non-linearity of the MFG utility function in the population distribution, which we highlight for MFGs.
We test our primary iterative EGTA framework in MFGs with various configurations and demonstrate the improved learning performance of EGTA over directly applying FP \citep{perrin2020fictitious} to MFGs.

Despite its effectiveness for solving MFGs, the primary method requires a large number of utility simulations due to the non-linearity of the utility function in the population distribution, which we refer to as low sample efficiency in EGTA.
To improve the sample efficiency, we introduce a GML approach and a regularized subroutine to our iterative EGTA framework.
The GML approach is a form of regression that learns the utility function progressively over EGTA iterations.
With a learned utility function, utility information can be predicted, and thus reducing the number of queries to the simulator.
The regularized subroutine improves sample efficiency through reducing the number of iterations of subroutines in each EGTA iteration.
In a separate work \cite{wang2023regularization}, we have shown that properly regularizing the best response target (i.e., not best-responding to an exact equilibrium) will lead to an improved learning performance for EGTA and the regularization can be achieved by early stopping a subroutine within each EGTA iteration.
For our purposes, early stopping a subroutine means less utility queries for each iteration of EGTA and improved sample efficiency if the overall learning performance would not decline.
By introducing GML and regularization, we demonstrate a significant improvement on the sample efficiency (i.e., EGTA with GML and regularization only requires $1/6$ of simulations needed by the primary method) over a variety of MFG configurations.

The theoretical results of this work are twofold.
First, we prove the existence of NE in an empirical MFG with a restricted strategy space under a mild assumption, assuming the MFG is fully symmetric.
Second, we prove that the iterative EGTA converges to NE of the full game if the best response target is NE across iterations and an exact best-response oracle is available.

Contributions of this study include: 
\begin{enumerate}
    \item We propose an iterative EGTA framework for solving MFGs with FP and RD as two practical subroutines for the empirical game analysis. 
    \item By introducing GML and regularization, we significantly improve the sample efficiency of the iterative EGTA framework. 
    \item We demonstrate the effectiveness of approaches for solving MFGs in various configurations.
    \item Theoretically, we prove the existence of NE in an empirical MFG among a restricted set of strategies under a mild assumption and prove the convergence of the iterative EGTA to NE of the full MFG.

\end{enumerate}

\section{Related Work}
\subsection{Learning Mean Field Games}
There is a large prior literature on learning solutions of MFGs. 
Here we survey few that are closely related to our approach.
\citet{elie2020convergence} first studied the convergence of approximate discrete-time FP in MFGs.
\citet{perrin2020fictitious} further proved the convergence rate of continuous-time FP in MFGs and extended the study to MFGs with common noise.
Moreover, FP for MFGs with finite time horizons was demonstrated effective in their work.
\citet{perolat2021scaling} criticized that FP is not scalable to MFGs with large state spaces and proposed Online Mirror Descent (OMD) as a solution.
They empirically showed that OMD converges significantly faster than FP in MFGs with a large number of states.
In the aforementioned algorithms, a fixed initial distribution of the population is required.
\citet{perrin2021generalization} argued that a fixed initial distribution restricts the practical applications of MFGs since a real initial distribution could be different from the one used for training.
They proposed a learning algorithm to learn a \textit{Master Policy}, which takes the distribution of population as input and thus taking the initial distribution into consideration.
They demonstrated the ability of generalization of the learned master policy.

In a recent work that is simultaneous with and independent of our \textbf{primary method}, \citet{muller2021learning} adapted PSRO to MFGs and analyzed convergence properties with various solution concepts.
Both theirs and our primary method are based on an iterative EGTA/PSRO framework for MFGs, highlighting the issue that the utility function for MFGs is not generally linear in the distribution.
Theoretically, both works prove the existence of NE in the empirical game with restricted strategy set and the convergence of iterative EGTA/PSRO to NE.
What is unique to their work is that they investigate the modifications for PSRO to converge to (coarse) correlated equilibria in MFGs as well as the theoretical counterparts.
The contribution of our primary method focuses on practical techniques for the empirical game analysis and goes beyond their work in including full details of how online learning algorithms (i.e., FP and RD as subroutines) realize the framework of EGTA for MFGs. 
We also include more experimental results on performance of these methods for deriving approximate equilibria for MFGs.
Since both theirs and our primary method rely on either black-box optimization or online approaches for computing NE of intermediate empirical games, the issue of low sample efficiency exists. 
As a major contribution of our work, our GML and regularization successfully addresses this issue.

\subsection{Game Model Learning}

To the best of our knowledge, our work is the first study that combines GML with the iterative EGTA framework.
Here we briefly survey prior work on GML that assumes the full game is given.
\citet{Vorobeychik07} first introduced the concept of learning normal-form game models as utility function regression from simulation data sampled over continuous strategy spaces.
They demonstrated the approach using single-parameter strategy representation.
\citet{Ficici08} clustered a large number of players into two roles based on data consisting of strategy profiles and utilities. 
Then regression of the utility function was applied for each role.
\citet{wiedenbeck2018regression} deployed Gaussian process regression to learn the utility function of large symmetric games.
\citet{sokota2019learning} extended GML to role-symmetric games by regressing the deviation payoff function rather than the payoff function.

\section{Preliminaries}
We study MFGs with a temporal structure, where representative players interact with populations over a shared state space $X$, action space $A$ and time horizon $T$. 
A \textit{multi-population mean field game} (MP-MFG) in normal form is given by $\mathcal{G}=([N_p],(S_i),(u_i))$.
$[N_p]=\{1,\dotsc,N_p\} $ is a set of $N_p$ populations indexed by~$i$.
Each population corresponds to a conceptually infinite and interchangeable set of players playing a particular role in the game.
$S_i$ denotes the set of strategies of population $i$ and $u_i$ denotes the utility function.
State space $X$, action space $A$ and time horizon $T$ are encoded in the definitions of a strategy $s_i\in S_i$ and the utility function $u_i$ as described below.


A strategy of the representative player of population $i$ at time $t$, $s_{i,t}$, maps from state space $X$ to the space of action distributions $\Delta(A)$, where $\Delta$ represents the probability simplex over a set.
The overall strategy $s_i=(s_{i,t})_{t\in[0, T-1]} \in S_i$ is a sequence of strategies from time~0 through horizon $T$\@.

The behavior of each population is summarized by a mean field, defined as distributions over an underlying state space $X$ of the game environment.
Denote distributions of all populations over time horizon $T$ as $\mu=\{\mu_1,\dotsc,\mu_{N_p}\}$, where $\mu_i = (\mu_{i,t})_{t\in[0, T]} \in \Delta(X)^{T+1}$.

Utility functions $u_i: S_i \times \Delta(X)^{N_p\times (T+1)} \rightarrow \mathbb{R}$ define the payoff to a representative player of population $i\in [N_p]$ playing its strategy against the distributions of all populations $\mu$ (i.e., mean fields).

A mixed strategy $\sigma_i$ is a probability distribution over strategies in $S_i$, with $\sigma_i(s_i)$ denoting the probability the representative player of population~$i$ plays strategy $s_i$. 
Let $\sigma$ be the profile of strategies across populations (i.e., $\sigma=(\sigma_1,\dotsc, \sigma_{N_p})$).

The representative player of population~$i$'s \textit{best response} to population distributions $\mu$ is any strategy yielding maximum payoff for the player, holding the distributions $\mu$ constant:
\begin{displaymath}
\br_i(\mu) = \argmax_{\sigma_i\in \Delta(S_i)} u_i(\sigma_i, \mu).
\end{displaymath}
Let $\br(\mu)=\prod_{i \in N_p}\br_i(\mu)$ be the overall best-response correspondence for populations’ distributions $\mu$. 
Here $\prod$ is the Cartesian product. 
A Nash equilibrium for an MFG is a profile $\sigma^*$ such that $\sigma^*\in \br(\mu^*)$, where $\mu^*$ is induced by $\sigma^*$. 

A distribution $\mu_i$ is said to be induced by $\sigma_i$, denoted as $\mu^{\sigma_i}_i$, following the \textit{Forward Equation}, that is, given initial distribution $\mu_{i,0}$, for $t\in[0,T-1]$ and all $x'_i \in X$,
\begin{equation}
    \mu_{i,t+1}^{\sigma_{i}}(x'_i) = \sum\limits_{x_i,a_i\in X,A} \mu_{i,t}^{\sigma_{i}}(x_i) \sigma_{i,t}(a_i\mid x_i) p(x'_i\mid x_i,a_i),
\label{eq:forward}
\end{equation}
where $p:X\times A\rightarrow \Delta X$ is the transition function.

The representative player of population~$i$'s \textit{regret} in profile $\sigma$ given distributions $\mu$ in game $\mathcal{G}$ is given by 
\begin{displaymath}
    \rho^{\mathcal{G}}_i(\sigma_i, \mu) = \max_{s_i\in S_i}u_i(s_i, \mu)-u_i(\sigma_i, \mu).
\end{displaymath}
Regret captures the maximum the representative player of population~$i$ can gain in expectation by unilaterally deviating from its mixed strategy in $\sigma$ to an alternative strategy in $S_i$\@, given distributions $\mu$. 
A NE strategy profile has zero regret for each representative player. 
A profile is said to be an $\epsilon$-Nash equilibrium ($\epsilon$-NE) if no representative player can gain more than $\epsilon$ by unilateral deviation. 
The regret of a strategy profile $\sigma$ is defined as the sum over representative players' regrets:%
\begin{displaymath}
    \rho^{\mathcal{G}}(\sigma, \mu) = \sum_{i\in [N_p]} \rho^{\mathcal{G}}_i(\sigma_i, \mu).
\end{displaymath}

An \textit{empirical MFG} $\mathcal{G}_{S\downarrow \Lambda}$ is an approximation of the true game $\mathcal{G}$, in which populations choose from a restricted strategy set $\Lambda_i\subseteq S_i$, and payoffs are estimated through simulation. 
That is, $\mathcal{G}_{S\downarrow \Lambda} =([N_p],(\Lambda_i),(\hat{u}_i))$, where $\hat{u}$ is a projection of $u$ onto the restricted strategy space $\Lambda$\@.%

\section{Iterative EGTA for MFGs}

\subsection{Framework}
In finite games, an iterative approach for computing NE is presented by the well-known Double Oracle (DO) algorithm \citep{mcmahan2003planning}, which alternates between analysis of a current empirical game and adding strategies that best respond to a current equilibrium. 
We extend DO to MFGs and provide a game-solving algorithm for MP-MFGs in Algorithm~\ref{alg:DO MP-MPG} as our primary EGTA method for MFGs. 

In Algorithm~\ref{alg:DO MP-MPG}, for each population $i$, the representative player is initialized with a strategy $s_{i,0}$ and an initial distribution $\mu_{i,0}$.
At each iteration $\tau$, an equilibrium $\sigma^e$ of the current empirical game is computed by a subroutine (e.g., FP or RD) and then distributions $\mu^e$ are induced by $\sigma^e$ through the forward equation. 
Then for each population $i$, the representative player computes an exact/approximate best response strategy $s_{i,\tau}$ to the distribution $\mu^e$ and adds it to the corresponding strategy set in the empirical game.
This process repeats until no beneficial deviation strategy could be found or certain stopping criterion is satisfied.

\begin{algorithm}[!ht]
\caption{Iterative EGTA for MP-MPGs (primary method)}
\label{alg:DO MP-MPG}
\begin{algorithmic}[1] 
\REQUIRE an initial strategy $s_{i,0}$ and an initial distribution $\mu_{i,0}$ for all population $i\in [N_p]$.\\
\FOR{EGTA iteration $\tau \in \{1, \dotsc, \mathcal{T}\}$}
\STATE Compute $(\sigma^e, \mu^e)$ of the current empirical game by a subroutine
\FOR{$i \in \{1,\dotsc, N_p\}$}
\STATE Compute a best response strategy $s_{i,\tau} \in S_i$ to the empirical equilibrium distribution $\mu^e$
\STATE Add $s_{i,\tau}$ to the strategy set of population $i$: $\Lambda_i \gets \Lambda_i \bigcup s_{i,\tau}$
\ENDFOR
\ENDFOR

\STATE \textbf{Return} $(\sigma^e, \mu^e)$
\end{algorithmic}
\end{algorithm}

\subsection{Analyzing an Empirical MFG}\label{sec: Empirical Game Analysis}
In EGTA, analyzing an intermediate empirical game is crucial for generating effective strategies.%
\footnote{For discussion simplicity, we assume that players are fully symmetric (i.e., $N_p=1$) for illustrating the representation issue. So the population index $i$ is dropped when the context is clear. The analysis can be easily extended to MP-MFGs.}
Despite an MFG model reduces the analysis to the study of interactions between two parties (i.e., a representative player and the population), computing a NE of an empirical MFG is more than constructing an explicit payoff matrix and then applying a game-solver for two parties as in finite games.
Proposition~\ref{lemma: fake NE} shows that MFGs with a restricted strategy set can only be solved with an explicit payoff matrix under certain restrictive conditions.

Consider an empirical MFG with an explicit payoff matrix representation shown in Table~\ref{tab: single-population matrix}. 
In the empirical game, there are 4 strategies $\Lambda=\{s_0,s_1,s_2,s_3\}$ in the restricted strategy set.
Since the game is fully symmetric, the population would act following distributions $\mu=\{\mu_0,\mu_1,\mu_2,\mu_3\}$ induced by corresponding strategies in $\Lambda$.
In the payoff matrix, the value in entry $(j, k)$, where $j,k \in \{0,1,2,3\}$, is $u(s_j, \mu_k)$, the utility of the representative player playing $s_j$ against the distributions $\mu_k$ induced by $s_k$.

\begin{table}[!h]
    \centering
    \begin{tabular}{ |c|c|c|c|c| } 
 \hline
  & $s_0$ & $s_1$ & $s_2$ & $s_3$ \\ 
 \hline
 $s_0$ & $u(s_0,\mu_0)$ &$u(s_0,\mu_1)$  & $u(s_0,\mu_2)$ &$u(s_0,\mu_3)$ \\ 
 \hline
 $s_1$ & $u(s_1,\mu_0)$ &$u(s_1,\mu_1)$  & $u(s_1,\mu_2)$ &$u(s_1,\mu_3)$\\ 
 \hline
 $s_2$ & $u(s_2,\mu_0)$ &$u(s_2,\mu_1)$  & $u(s_2,\mu_2)$ &$u(s_2,\mu_3)$\\ 
 \hline
 $s_3$ & $u(s_3,\mu_0)$ &$u(s_3,\mu_1)$  & $u(s_3,\mu_2)$ &$u(s_3,\mu_3)$\\ 
 \hline
\end{tabular}
\caption{Single-population payoff matrix.}
\label{tab: single-population matrix}
\end{table}{}

\begin{proposition}\label{lemma: fake NE}
The NE of the aforementioned payoff matrix will not generally be a NE of the empirical mean field game unless the utility function is linear in $\mu$.
\end{proposition}
\begin{proof}
Assume $\sigma$ is a NE computed from the payoff matrix. 
According to the definition of NE, we have 
\begin{equation}\label{eq: matrix NE}
    \sum_{j\in[\abs{\Lambda}]} \sum_{k\in[\abs{\Lambda}]} \sigma(s_j)\sigma(s_k)u(s_j, \mu_k) \ge \sum_{k\in[\abs{\Lambda}]}\sigma(s_k)u(s', \mu_k), \forall s'\in \Lambda.
\end{equation}

According to the definition of NE $\sigma^*$ in MFG, we have 
\begin{equation}\label{eq: MFG NE}
    \sum_{j\in[\abs{\Lambda}]} \sigma^*(s_j)u(s_j, \mu^{\sigma^*}) \ge 
    u(s', \mu^{\sigma^*}),  \forall s'\in \Lambda
\end{equation}
where $\mu^{\sigma^*}$ is induced by ${\sigma^*}$.

By comparing inequalities~\ref{eq: matrix NE} and~\ref{eq: MFG NE}, to make the NE $\sigma$ an MFG NE $\sigma^*$, the following condition should hold
\begin{displaymath}
    \sum_{k\in[\abs{\Lambda}]} \sigma(s_k)u(s_j, \mu_k) = u(s_j, \mu^{\sigma}), \forall s_j\in \Lambda
\end{displaymath}
indicating the requirement of linearity in $\mu$ at least at the equilibrium point. 

\end{proof}

Since the utility function in MFGs will not generally be linear in $\mu$, analysis of an empirical game based on an explicit payoff matrix is impractical.
Instead, we rely on query-based methods.
We propose FP and RD as two subroutines for solving empirical games and query utility information through simulation as needed.

\subsubsection{FP for Empirical MFGs}
FP for MFGs has been studied by \citet{elie2020convergence} and \citet{perrin2020fictitious}.
We adapt it to analyzing an empirical game with a restricted set of strategies.
In Algorithm~\ref{alg: FP inner loop}, we demonstrate how to apply FP to an empirical game.
Specifically, starting from the uniform strategy $\Bar{\sigma}$ over the strategies in the restricted strategy set and its induced distribution $\Bar{\mu}$, at each iteration $j\in [1, J]$, the representative player of a population $i$ computes a best response strategy $s_{i,j}\in\Lambda_i$ against populations $\Bar{\mu}$. 
The probability in the mixed strategy of playing $s_{i,j}$ is updated by the frequency of $s_{i,j}$ appearing as a best response, that is, the count of being a best response $n_{s_{i,j}}$ divided by the total count up to the $j^{th}$ iteration. 
Mathematically, for all $i\in[N_p]$ and $k\in [|\Lambda_i|]$, 
the count of $s_{i,k}$ being a best response $n_{s_{i,k}}$ is incremented by 1 if $s_{i,k}$ is the best response at the current iteration (i.e., $s_{i,k} = s_{i,j}$) and remains the same otherwise. 
\[ n_{s_{i,k}} =\begin{cases} 
      n_{s_{i,k}} + 1 & s_{i,k} = s_{i,j} \\
      n_{s_{i,k}} & s_{i,k} \ne s_{i,j} \\
   \end{cases}
\]
Then we update the probability of every strategy in the mixed strategy by
\begin{equation}\label{eq: FP update}
    \Bar{\sigma}_i(s_{i,k})=\frac{n_{s_{i,k}}}{|\Lambda_i|+\sum_{k\in[|\Lambda_i|]}n_{s_{i,k}} }, \forall s_i\in \Lambda_i, i\in [N_p].
\end{equation}

\begin{algorithm}[!htpb] 
\caption{Fictitious Play as a Subroutine}
\label{alg: FP inner loop}
\begin{algorithmic}[1] 
\REQUIRE an empirical game. An initial averaged strategy $\Bar{\sigma}_i$ over strategies in the restricted set $\Lambda_i = (s_{i,1}, \dotsc ,s_{i, \tau})$ of population $i, \forall i \in [N_p]$ and distributions $\Bar{\mu}$ that are induced by $\Bar{\sigma}$.\\
\FOR{FP iteration $j \in \{1, \dotsc, J\}$}
\FOR{$i \in \{1,\dotsc, N_p\}$}
\STATE Compute a best response strategy $s_{i,j} \in \Lambda_i$ against $\Bar{\mu}$
\STATE $n_{s_{i,j}} \gets n_{s_{i,j}} + 1$
\ENDFOR
\FOR{$i \in \{1,\dotsc, N_p\}$}
\STATE Update $\Bar{\sigma}_i$ according to Equation~\ref{eq: FP update} and induce $\Bar{\mu}_i$
\ENDFOR
\ENDFOR

\STATE \textbf{Return} $(\Bar{\sigma}, \Bar{\mu})$

\end{algorithmic}
\end{algorithm}

To induce a corresponding distribution $\Bar{\mu}_i$, we first build a weighted average strategy that is equivalent to the mixed strategy $\Bar{\sigma}_i$. 
Specifically, consider a mixed strategy $\sigma\in \Delta(\Lambda)$ defined on the empirical game with a restricted strategy set $\Lambda$. An equivalent strategy $\Bar{s}$ is defined as, for each population $i$,
\begin{displaymath}
    \Bar{s}_{i,t}(a\mid x) = \frac{\sum\limits_{k=1}^{|\Lambda_i|}\sigma_i(s_{i,k,t})\mu_t^{s_{i,k,t}}(x) s_{i,k,t}(a\mid x)}{\sum\limits_{k=1}^{|\Lambda_i|}\sigma_i(s_{i,k,t})\mu_t^{s_{i,k,t}}(x)}, \forall t\in[0,T-1],
\end{displaymath}
where $s_{i,k,t}$ is the $k$\textsuperscript{th} strategy of population $i$ at time step $t$.
Then the induced distribution is computed through Equation~\ref{eq:forward} or estimated through various approaches (e.g., empirical density estimation \citep{perrin2020fictitious} or generative models \citep{perrin2021mean}).

The empirical game analysis stops until certain stopping criterion is satisfied (e.g., reaching a fixed number of iterations or the difference between the averaged strategies in two consecutive iterations is tiny).
FP has been proved for convergence to NE in MFGs with a single population \citep{elie2020convergence,perrin2020fictitious}.
However, FP will not generally converge to a NE (or even a CE) in MFGs with multiple populations.

\subsubsection{RD for Empirical MFGs}
Replicator dynamics describes an evolving trajectory of mixed profiles
and is commonly employed as a heuristic equilibrium search algorithm in finite games \citep{taylor1978evolutionary,smith1973logic}. 
In Algorithm~\ref{alg: RD}, we adapt RD to our MFG model and propose it as a practical subroutine for empirical game analysis.
Similar to RD in finite games, at each iteration the update of a strategy's probability in population $i$ is in proportion to the deviation payoff of that strategy from the average fitness, weighted by its probability from the previous iteration and a learning rate. 
Theoretically, RD has not been proved for convergence as FP. 
However, we show that RD exhibits empirical convergence with an even more stable learning manner than FP in our experiments.

\begin{algorithm}[!htpb] 
\caption{Replicator Dynamics as a Subroutine}
\label{alg: RD}
\begin{algorithmic}[1] 
\REQUIRE an empirical game. An initial averaged strategy $\Bar{\sigma}_i$ over strategies in the restricted set $\Lambda_i = (s_{i,1}, \dotsc ,s_{i, \tau})$ of population $i, \forall i\in [N_p]$ and $\Bar{\mu}$ is induced by $\Bar{\sigma}$. A learning rate $dt$.\\
\FOR{RD iteration $j \in \{1, \dotsc, J\}$}
\FOR{$i \in \{1,\dotsc, N_p\}$}
\STATE Compute the average fitness $F_i = u_i(\Bar{\sigma}_i, \Bar{\mu})$
\FOR{$s_i \in \Lambda_i$}
\STATE Evaluate $u_i(s_i, \Bar{\mu})$
\STATE Update $\Bar{\sigma}'_i(s_i) \gets \Bar{\sigma}_i(s_i) + dt*\Bar{\sigma}_i(s_i)[u_i(s_i, \Bar{\mu}) - F_i]$
\ENDFOR
\ENDFOR
\STATE $\Bar{\sigma} = \Bar{\sigma}'$
\STATE Induce new distribution $\Bar{\mu}$ using updated $\Bar{\sigma}$.
\ENDFOR
\STATE \textbf{Return} $(\Bar{\sigma}, \Bar{\mu})$

\end{algorithmic}
\end{algorithm}



\subsection{Convergence to NE}
We analyze the theoretical properties of iterative EGTA for MFGs from two aspects: the existence of NE in mean field empirical games and the convergence of iterative ETGA to NE of the full game. 
We assume that players are fully symmetric (i.e., a single population) and prove that the NE exists in an empirical game under one mild assumption.

\begin{assumption}\label{asmpt: continuity}
The utility function $u(s,\mu)$ is continuous in the distribution $\mu$.
\end{assumption}

\begin{theorem}\label{thm: existence of empirical NE}
Under Assumption~\ref{asmpt: continuity}, for single-population MFGs with finite state and action spaces, there exists a Nash equilibrium in an empirical game. 
\end{theorem} 

According to Theorem~\ref{thm: existence of empirical NE}, an empirical NE exists and hence it can be obtained through some theoretically proven equilibrium search subroutines (e.g., FP).
The remaining problem is whether iterative EGTA converges to the NE of the full game. 

\begin{theorem}\label{thm: DO convergence}
For single-population MFGs with finite state and action spaces, suppose the empirical NE is the best response target at each iteration of iterative EGTA and an exact best response oracle is available, then the empirical NE converges to the NE of the full game.
\end{theorem} 

We provide the proofs of both theorems in Appendix~\ref{app: Convergence}. 



\section{Game Model
Learning and Regularization for Improved Sample Efficiency}

One crucial factor that affects the practicality of our primary iterative EGTA method is sample efficiency.
Sample efficiency here refers to the number of simulations needed for estimating the utilities in the equilibrium search at each iteration of EGTA. 
In finite games, the utilities of any mixed strategy profile can be computed by taking an expectation of utilities across pure strategy profiles in the support, so utility information of pure strategy profiles can be re-used.

Unlike finite games, Proposition~\ref{lemma: fake NE} shows that the utility $u(s, \mu^\sigma)$ of playing strategy $s$ against population distributions $\mu^\sigma$ needs to be evaluated for different $\mu^\sigma$ and generally cannot be computed by first computing $u(s, \mu^{s'})$ for each pure strategy $s'$ in the support of $\sigma$ and then taking an expectation as in finite games.
Moreover, since it is very unlikely to encounter a same distribution $\mu^\sigma$ across EGTA iterations, it is also not useful to store the corresponding utility information. 
Due to these characteristics, instead of storing utility information and re-using them, the iterative EGTA approaches up to this point (including our primary method and the version by \citet{muller2021learning}) need to compute or simulate $u(s, \mu^\sigma)$ whenever $\mu^\sigma$ changes in the equilibrium computation (i.e., query-based implementation), which results in a low sample efficiency.
To improve the sample efficiency, we introduce GML and regularization to our primary method. 
For discussion purposes in this section, we assume that MFGs are single-population (i.e., $N=1$) and our approach can be readily extended to MFGs with multiple populations.

\subsection{Game Model Learning} \label{sec: GML}
Our GML approach is a form of regression that learns the utility function based on utility information collected over previous EGTA iterations.
With a game model (i.e., a learned utility function), sample efficiency can be improved by querying the game model rather than running simulations as long as the game model is able to provide high-fidelity predictions on these queries.
In the following discussion, we first discuss where a game model fits in the EGTA framework and then elaborate the method for learning a game model given utility samples.

\subsubsection{Applying GML to EGTA}
To implement GML in iterative EGTA, one key step is to periodically update the game model based on collected utilities and apply the game model to running a subroutine for equilibrium computation of the current empirical game.
We select RD as a subroutine for illustration purposes.

\begin{figure}[!htpb]
\centering
\includegraphics[width=0.7\columnwidth]{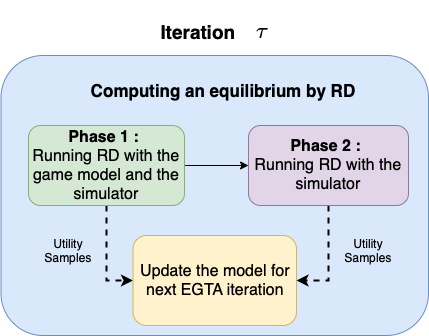}
\caption{RD with a game model.}\label{fig: framework}
\end{figure}

Denote the current EGTA iteration as iteration $\tau$. 
Our object is to approximate the NE of the current empirical game with the strategy set $\Lambda_\tau$, using RD and the game model $\hat{u}_{\tau-1}$ learned based on $\Lambda_{\tau-1}$ (i.e., the game model learned from previous iterations).
Since the model $\hat{u}_{\tau-1}(s, \mu^\sigma)$ only contains utility information of $s\in \Lambda_{\tau-1}$ and $\sigma \in \Delta\Lambda_{\tau-1}$ from previous iterations, while RD requires utilities $u(s, \mu^\sigma), \forall s\in \Lambda_{\tau}, \forall \sigma \in \Delta\Lambda_{\tau}$ for the current iteration, we cannot directly apply the model $\hat{u}_{\tau-1}$ due to the lack of information of $s_\tau$.
To handle this issue, we interleave utility approximation with simulations for different strategies.
In particular, for $s\in \Lambda_{\tau-1}$ and $\sigma \in \Delta\Lambda_{\tau}$, we project $\sigma$ onto $\Delta\Lambda_{\tau-1}$ by $P_{\tau - 1}(\sigma) = \argmin_{\sigma'\in \Delta\Lambda_{\tau-1}} ||\sigma' - \sigma ||_2$ and approximate $u(s, \mu^\sigma)$ by $\hat{u}_{\tau-1}(s, \mu^{P(\sigma)})$.
Our assumption here is that if $P(\sigma)$ is close to $\sigma$, then $u_{\tau-1}(s, \mu^\sigma)$ is close to $u_{\tau-1}(s, \mu^{P(\sigma)})$ and the model is valid for estimating $u(s, \mu^\sigma)$.
The scenario in the assumption often holds at the start of running RD when RD is initialized with the equilibrium strategy from last iteration.
Since the update of strategy in RD is controlled by a small step size, $P(\sigma)$ will be close to $\sigma$ within first few RD iterations.
For $s = s_\tau$, we query the simulator (e.g., a noiseless simulator) to obtain the exact utility $u(s, \mu^\sigma)$ because the game model does not contain its utility information.
We refer to this procedure as the first phase.

The number of RD iterations for the first phase is determined by the quality of utility estimations given by the projection.
To measure this quality, we set a threshold for the L-2 distance between $\sigma$ and its projection $P(\sigma)$.
If the distance goes beyond the threshold, it means that the game model with projection becomes less accurate on the utility predictions.
So we should stop using the model and switch to the simulator (i.e., the second phase).  

In the second phase, we directly run RD with the simulator for all $s\in \Lambda_{\tau}$ and $\sigma \in \Delta\Lambda_{\tau}$ for a fixed number of iterations. 
At the end the second phase, utilities collected from the simulator at both phases are used to fine-tune the current game model. 
Note that the sampling of these utilities is guided by RD to avoid sampling the whole strategy space, where the latter will hurt our motivation of improving sample efficiency.
The overall framework is depicted in Figure~\ref{fig: framework}.


\subsubsection{Learning Utility Functions}
In the previous section, we take a game model from previous EGTA iterations for granted, and now we describe how the model is updated based on utility samples.

Given an empirical game at iteration $\tau$, we aim at learning a game model (i.e., a utility function) $\hat{u}_{\tau}(\sigma, \mu^{\sigma'}), \forall \sigma, \sigma' \in \Delta\Lambda_{\tau}$ given the model from previous iterations $\hat{u}_{\tau-1}$ (which can be a random function for $\tau=1$) and new utility data generated at the current iteration $\tau$.
To obtain the utility $\hat{u}(\sigma, \mu^{\sigma'})$ defined on mixed strategies $\sigma\in \Delta \Lambda_{\tau}$, it is sufficient to learn a game model $\hat{u}(s, \mu^{\sigma'})$ on pure strategies $s\in \Lambda_\tau$ and take an expectation.
We focus on a general setup of MFGs where a strategy $s$ and distributions $\mu$ are generally both time-dependent (i.e., $s=(s_t)_{t\in [0,T-1]}$ and $\mu=(\mu_t)_{t\in [0,T]}$).
Explicitly encoding them as inputs to a learner (either a neural network or other regression models) will result in a high-dimensional input vector and entail an impractically complex regression setup.

To handle this issue, we apply a \textit{coarse coding} scheme from a separate work.
Specifically, let $I: \Lambda_\tau \rightarrow \mathbb{Z_+}$ be an indexing function that assigns a unique index to each strategy $s\in \Lambda$.
Let $\sigma \in \Delta \Lambda_\tau$ be the mixed strategy that induces the distribution $\mu^\sigma$.
Since the distribution induction function (Eq.~\ref{eq:forward}) is deterministic, the distributions $\mu^\sigma$ are uniquely determined by a fixed initial distribution $\mu_0 \in \Delta(X)$ and a strategy $\sigma$.%
\footnote{We assume the initial distributions $\mu_0$ are fixed, so it is sufficient for $\sigma$ to determine $\mu^\sigma$.}
Rather than learning $\hat{u}(s, \mu^\sigma)$ with time-dependent inputs, we learn a game model $\hat{u}: I(\Lambda_\tau) \times \Delta(\Lambda_\tau) \rightarrow \mathbb{R}$ with equivalent representations $I(s)$ and $\sigma$ of $s$ and $\mu^\sigma$.

\begin{figure}[!htpb]
\centering
\includegraphics[width=0.6\columnwidth]{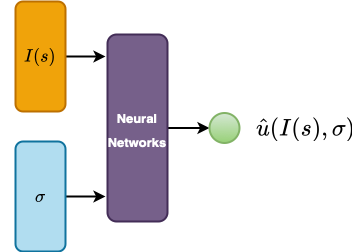}
\caption{Neural network structure for coarse coding.}\label{fig: coarse NN}
\end{figure}

Based on coarse coding, a utility data point is constructed to include an index of a pure strategy $I(s)$, a mixed strategy $\sigma$, and a utility target $u(s, \mu^{\sigma})$.
Since the object is to predict the true utility $u(s, \mu^\sigma)$ by $\hat{u}(I(s), \sigma)$, we fine-tune $\hat{u}_{\tau-1}$ by minimizing the mean square error $E[u(s, \mu^\sigma) - \hat{u}(I(s), \sigma))^2]$.
Our regression is based on neural networks and the structure of the neural networks is shown in Figure~\ref{fig: coarse NN}.

\begin{figure*}[!htpb]
\centering
    \subfloat[1-D(10, 10)]{\includegraphics[width=0.32\linewidth]{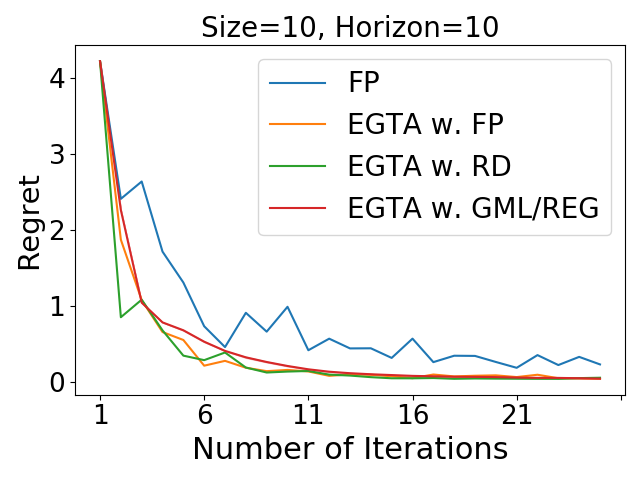}\label{fig:1d first}}\
    \subfloat[1-D(10, 20)]{\includegraphics[width=0.32\linewidth]{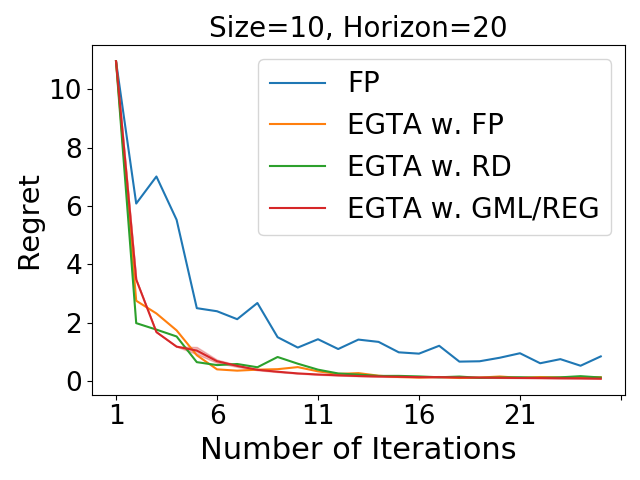}}\
    \subfloat[1-D(10, 30)]{\includegraphics[width=0.32\linewidth]{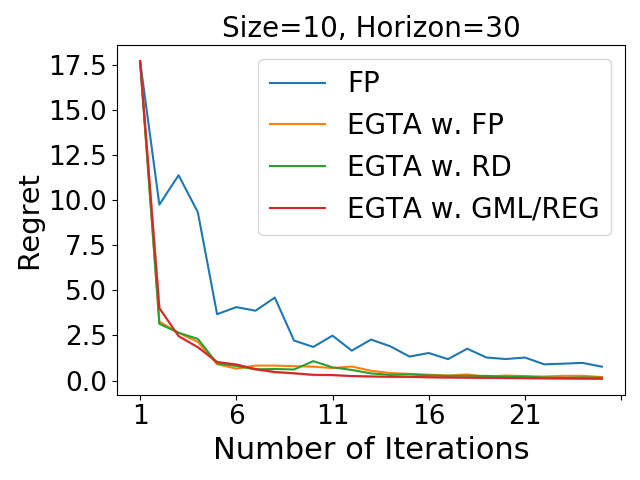}}\\
    \subfloat[1-D(50, 10)]{\includegraphics[width=0.32\linewidth]{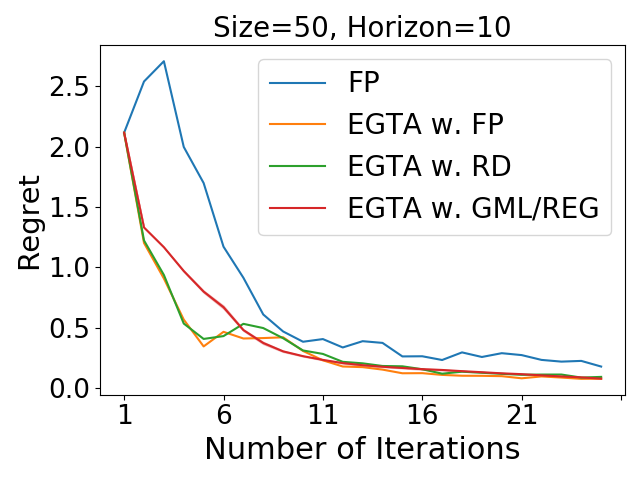}}\
    \subfloat[1-D(50, 20)]{\includegraphics[width=0.32\linewidth]{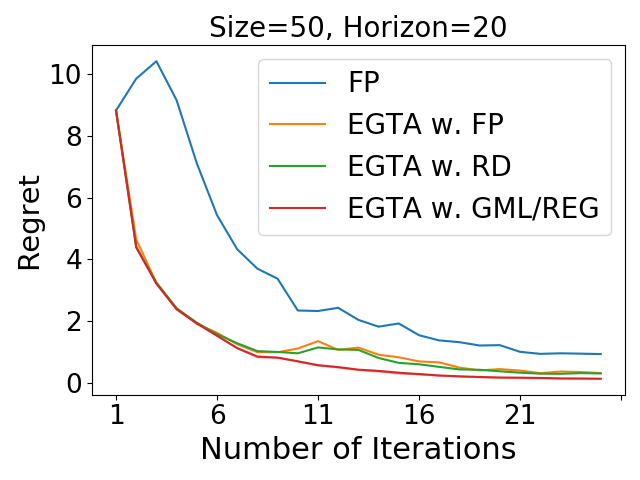}}\
    \subfloat[1-D(50, 30)]{\includegraphics[width=0.32\linewidth]{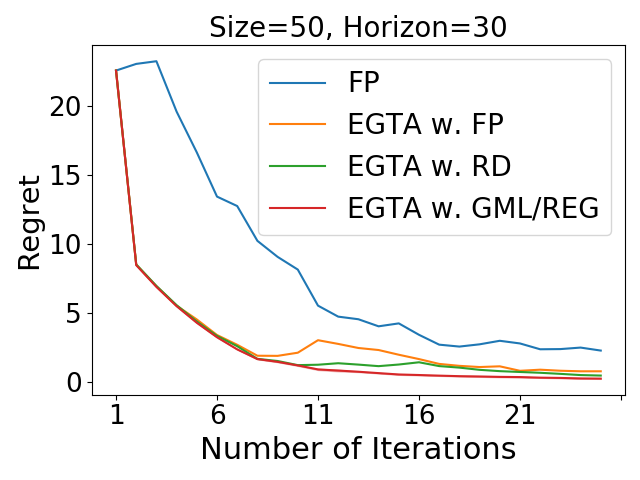}\label{fig:1d RD good}}\\
    \subfloat[1-D(100, 10)]{\includegraphics[width=0.32\linewidth]{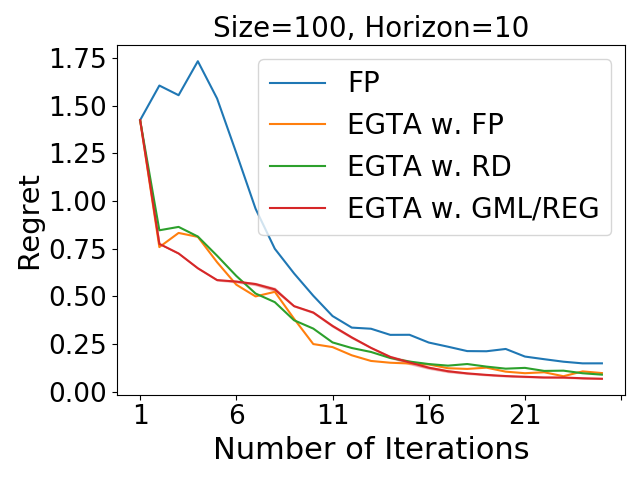}}\
    \subfloat[1-D(100, 20)]{\includegraphics[width=0.32\linewidth]{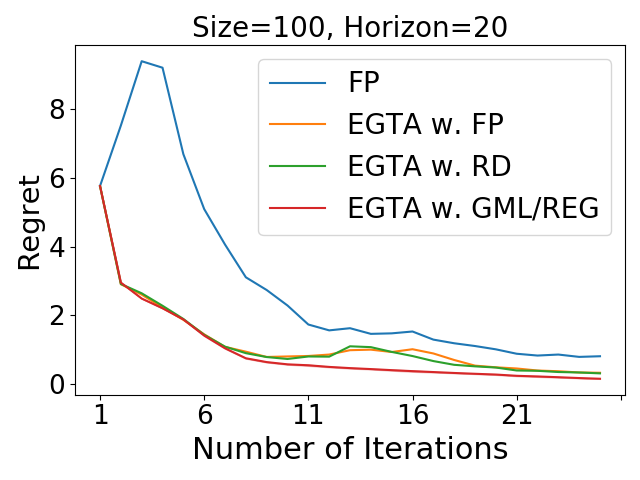}}\
    \subfloat[1-D(100, 30)]{\includegraphics[width=0.32\linewidth]{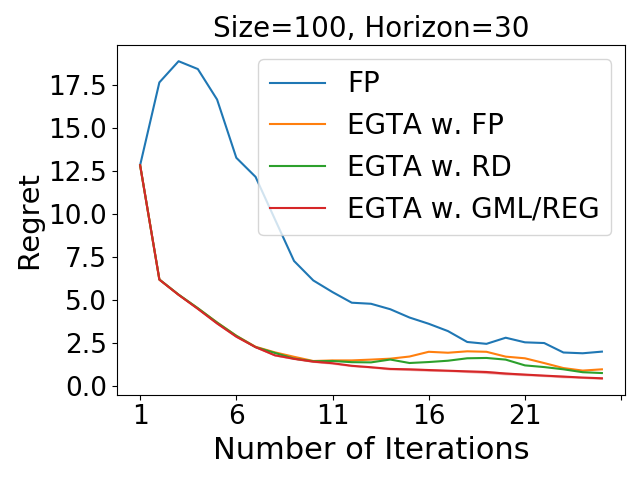}\label{fig:1d last}}\\
    \subfloat[2-D(100, 10)]{\includegraphics[width=0.32\linewidth]{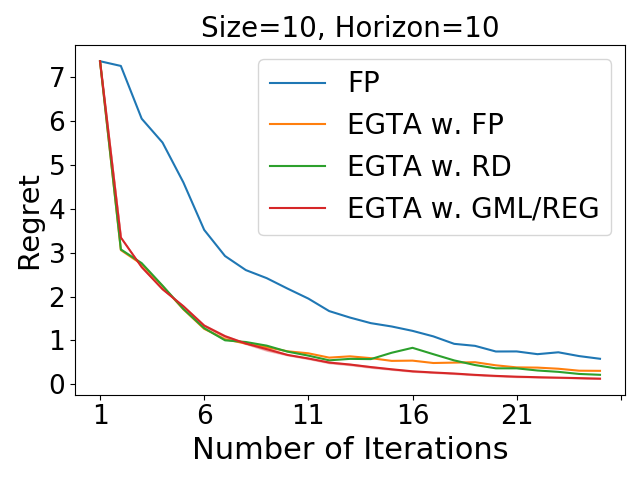}\label{fig:2d first}}\
    \subfloat[2-D(100, 20)]{\includegraphics[width=0.32\linewidth]{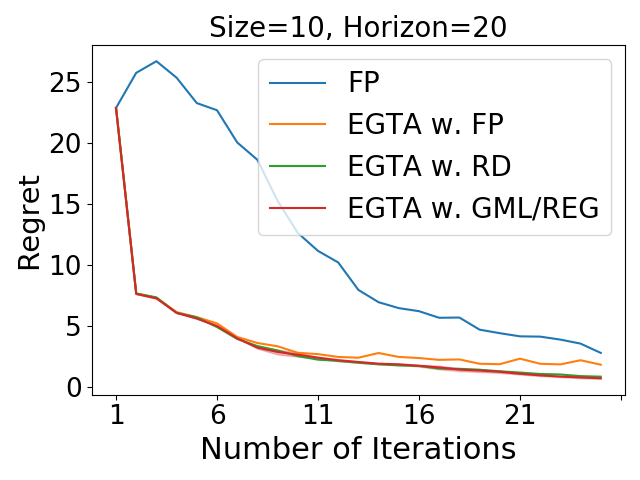}\label{fig:2d RD good}}\
    \subfloat[2-D(100, 30)]{\includegraphics[width=0.32\linewidth]{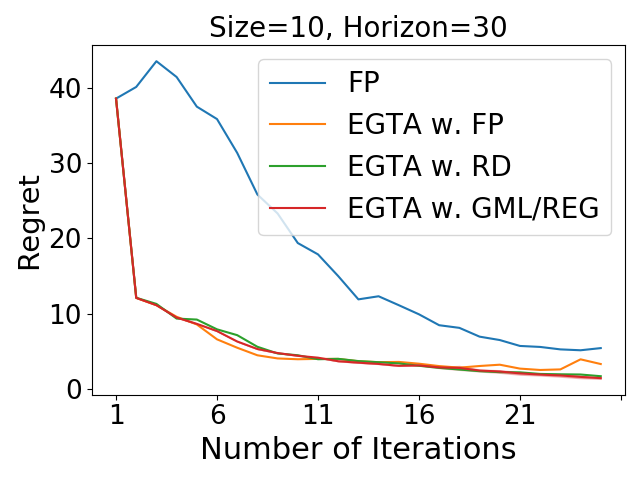}\label{fig:2d last}}\\
\caption{Experimental results of 1-D and 2-D beach bar problems.}\label{fig: 1_D Beach Bar}
\end{figure*}

\subsection{Regularization}
Our regularization method improves sample efficiency through reducing the number of iterations of RD in each EGTA iteration.
In current iterative EGTA approaches, the number of iterations for the subroutines (e.g., RD and FP) is set to be large enough so that they can approximately converge to the current equilibrium of the empirical game. 
This mimics the classic double oracle method, in which a best response to the current equilibrium is computed at each iteration.

In a separate work \cite{wang2023regularization}, we have shown that properly regularizing the best response target (i.e., not best-responding to an exact equilibrium) will lead to improved learning performance for EGTA. 
Regularization can be achieved by early stopping RD when the regret of the current profile with respect to the empirical game exceeds a pre-defined regret threshold.
For our purposes, early stopping of RD means less utility queries from the simulator and improved sample efficiency if the overall learning performance would not decline.
We apply this approach to our EGTA framework by replacing the regret threshold with a fixed number of RD iterations since the prediction error given by the game model could affect the regret estimation for reaching a threshold.

\subsection{Algorithms}
We show the complete EGTA framework with GML and regularization for single-population MFGs in Algorithm~\ref{alg: model EGTA} and Algorithm~\ref{alg: RD modeling learning}.
Compared to our primary method (Alg.\ref{alg:DO MP-MPG}), the main differences are the introduction of a game model $\hat{u}$ in Alg.~\ref{alg: model EGTA} and how the model $\hat{u}$ is updated and applied to RD in Alg.~\ref{alg: RD modeling learning} (discussed in Section~\ref{sec: GML}).
Note that, in Alg.~\ref{alg: RD modeling learning}, RD is initialized with the equilibrium $\sigma^e$ from last EGTA iteration, perturbed by a function $\delta$ to guarantee a full support (make sure every strategy can be played with non-zero probability and hence can be updated by RD).
Regularization is achieved by controlling the maximal number of RD iteration $J$.

\begin{algorithm}[!htbp]
\caption{Iterative EGTA with GML and regularization}
\label{alg: model EGTA}
\begin{algorithmic}[1] 
\REQUIRE An initial strategy $\Lambda_0= \{s_0\}$ and an initial distribution $\mu_0$. A neural network $\hat{u}$.\\
\STATE $\sigma^e \leftarrow s_0$
\STATE Initialize $\mu^e$ by Eq.~\ref{eq:forward} using $s_0$
\FOR{EGTA iteration $\tau \in \{1, \dotsc, \mathcal{T}\}$}
\STATE Compute a best response strategy $s_{\tau}$ to the empirical equilibrium distribution $\mu^e$
\STATE Add $s_{\tau}$ to the strategy set of population $i$: $\Lambda_\tau \gets \Lambda_{\tau-1} \bigcup s_{\tau}$
\STATE Compute $\sigma^e, \mu^e, \hat{u} \leftarrow \text{a subroutine } \Psi(\mathcal{G}_{S\downarrow \Lambda_\tau}, \hat{u}, \sigma^e, \mu^e)$  
\ENDFOR
\STATE \textbf{Return} $(\sigma^e, \mu^e)$
\end{algorithmic}
\end{algorithm}

\begin{algorithm}[!htpb] 
\caption{Regularized Replicator Dynamics as a Subroutine $\Psi$}
\label{alg: RD modeling learning}
\begin{algorithmic}[1] 
\REQUIRE An empirical game $\mathcal{G}_{S\downarrow \Lambda_\tau}$. A learned utility simulator $\hat{u}$. Equilibrium strategy $\sigma^e$, and distribution $\mu^e$.  \\ 

\textbf{Parameters:} A distance threshold $\gamma$. A maximal number of iterations for applying the model $M$. A learning rate $dt$.\\

\STATE Initialize a strategy $\Bar{\sigma}\leftarrow \delta(\sigma^e)$
\FOR{RD iteration $j \in \{1, \dotsc, J\}$}
\STATE $\Bar{\sigma}^p \leftarrow P_{\tau-1}(\Bar{\sigma})$
\IF{$||\Bar{\sigma}, \Bar{\sigma}^p||_2 < \gamma$ and $j < M$}
\STATE Approximate $u(s, \mu^{\Bar{\sigma}})$ by $\hat{u}(I(s), \Bar{\sigma}^p), \forall s \in \Lambda_{\tau-1}$
\STATE Simulate $u(s_{\tau}, \mu^{\Bar{\sigma}})$
\ELSE
\STATE Simulate $u(s, \mu^{\Bar{\sigma}}), \forall s \in \Lambda_\tau$
\ENDIF 

\STATE Save new data points $I(s)$, $\Bar{\sigma}$, and $u(s, \mu^{\Bar{\sigma}})$

\STATE Compute fitness $F = \sum_{s \in \Lambda}\Bar{\sigma}(s)u(s, \mu^{\Bar{\sigma}})$
\FOR{$s \in \Lambda_\tau$}
\STATE $\Bar{\sigma}(s) \leftarrow \Bar{\sigma}(s) + dt*\Bar{\sigma}(s)[u(s, \mu^{\Bar{\sigma}}) - F]$
\ENDFOR

\ENDFOR

\STATE Fine-tune $\hat{u}$ with all new data points
\STATE Compute the induced distributions $\mu^{\Bar{\sigma}}$ by Eq.~\ref{eq:forward} using $\Bar{\sigma}$
\STATE \textbf{Return} $\Bar{\sigma}$, $\mu^{\Bar{\sigma}}$, and $\hat{u}$

\end{algorithmic}
\end{algorithm}

\section{Experimental Results}

\subsection{The 1-D Beach Bar}
We consider a simplified version of Santa Fe bar problem \citep{arthur1994inductive,farago2002fair} and adopt the model by \citet{perrin2020fictitious}. 
Specifically, a beach bar problem for a single-population MFG is a Markov Decision Process with $|X|$ states on a one-dimensional
torus $(X ={0,\dotsc,\abs{X}-1})$.
Without loss of generality, we designate a bar to the state 0.
Positions of players are initialized according to a uniform distribution.
Players can keep still $(a_t=0)$ or move left $(a_t=-1)$ or right $(a_t=1)$ at time step $t$ on the torus to get as close as possible to the bar, while avoiding the crowded areas.
The transition function is given by:
\begin{displaymath}
    x_{t+1} = x_{t} + a_t + \epsilon_t
\end{displaymath}
where $a_t$ is an action of the representative player at time $t$ and $\epsilon_t$ represents the randomness of the environment.
The immediate reward function is given by:
\begin{displaymath}
    r(x_t, a_t, \mu_t) = \Tilde{r}(x_t) - \frac{|a_t|}{|X|} - \log(\mu_t(x_t))
\end{displaymath}
where $\Tilde{r}(x_t)$ measures the closeness to the bar from state $x_t$, $-\frac{|a_t|}{|X|}$ is the running cost and $-\log(\mu_t(x_t))$ represents the aversion of players to the crowded areas.


We test the performance of iterative EGTA in the 1-D beach bar problem with various configurations, which are determined by the Cartesian product of $|X|\in\{10, 50, 100\}$ and $T\in\{10,20,30\}$, denoted by 1-D($|X|$, $T$). 
In Figure~\ref{fig:1d first}-\ref{fig:1d last}, we plot the regret curves of iterative EGTA (with FP and RD as subroutines respectively) against directly applying FP to MFGs \citep{perrin2020fictitious}, where x-axis being the EGTA iterations.
Since the game size supports exact best response calculation and exact strategy evaluation given a fixed initialization and parameters of the randomness of the environment (i.e., the simulator is noiseless), the regret curves can be exactly computed and hence no error bar is reported in the plots.

\begin{figure}[!htpb]
\centering
\includegraphics[width=0.75\columnwidth]{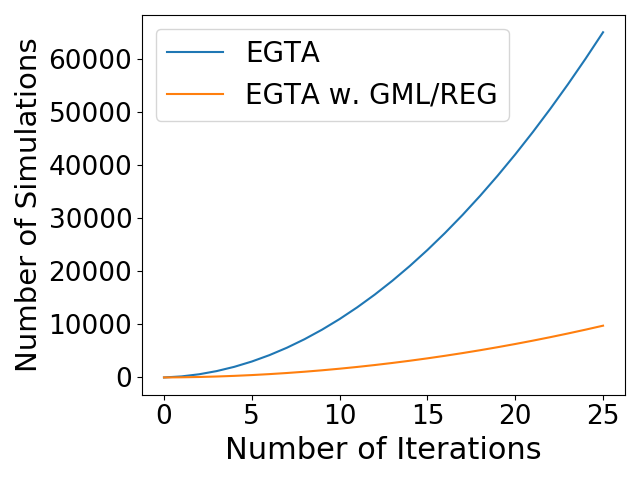}
\caption{The number of utility samples across EGTA iterations.}\label{fig: num samples}
\end{figure}

From Figure~\ref{fig:1d first}-\ref{fig:1d last}, we observe that the performance of our primary EGTA method (orange and green curves) dominates FP in all instances.
Moreover, as the instance becomes complex (i.e., with more states and longer horizon), the performance gap between our primary method and FP becomes more apparent. 
For the subroutine selection, we observe that RD in some cases exhibits a more stable learning manner (e.g., Fig~\ref{fig:1d RD good}) than FP while in most cases their performances are almost indistinguishable. 
For EGTA with GML and regularization (abbr. REG in the plots) (red curve), we observe that its performance also almost coincides with our primary method, but becomes even most stable as time horizon increases due to regularization.
Thanks to GML and regularization, we obtain this performance with only $1/6$ of the total utility queries compared to our primary method, which demonstrates a significant improvement on the sample efficiency.
We show the number of simulations needed across EGTA iterations in Figure~\ref{fig: num samples}.

\subsection{The 2-D Beach Bar}
We test the performance of our primary EGTA method in the 2-D beach bar problem with fixed $|X|=100$ and different time horizon $T\in\{10,20,30\}$, denoted by 2-D($|X|$, $T$). 
Compared to the 1-D beach bar problem, the 2-D beach bar problem includes extra actions (i.e., up, down, left, right, stay) yielding a more complex environment.
From Figure~\ref{fig:2d first}-\ref{fig:2d last}, we observe the same phenomenon as in the 1-D problem, that is, iterative EGTA dominates FP and as the game instance becomes larger, the advantage of iterative EGTA becomes apparent.
EGTA with GML and regularization again exhibits similar performance compared to the primary method but only consumes $1/6$ of the utility queries.

\subsection{Multi-population Chasing}
For MP-MFGs, we test the performance of iterative EGTA in a 3-population chasing problem \citep{perolat2021scaling} and observe that iterative EGTA also exhibits faster convergence than FP in the 3-population MFG.
We report the environment definition and our experimental results in Appendix~\ref{app: pred}.

\section{Conclusion}

We propose an iterative EGTA framework for computing NE in MFGs and a sample-efficient version by combining GML and regularization.
We demonstrate the efficacy of our approaches in various MFGs.
Theoretically, we prove the existence of NE in empirical MFGs and the convergence of the iterative EGTA framework.

\begin{acks}
This work was supported in part by funding from the US Army Research Office (MURI grant W911NF-18-1-0208), and a grant from the Effective Altruism Foundation.
\end{acks}

\balance
\bibliographystyle{ACM-Reference-Format}
\bibliography{thebib, wellman}

\clearpage
\appendix

\section{Convergence Analysis} \label{app: Convergence}

\subsection{Proof of Theorem~\ref{thm: existence of empirical NE}}
\begin{proof}[Proof of Theorem~\ref{thm: existence of empirical NE}]
To prove the existence of a NE in the empirical game with strategy sets $\Lambda$ using Kakutani's fixed point theorem \citep{kakutani1941generalization}, we need to show
\begin{enumerate}
    \item The empirical strategy space $\Delta(\Lambda)$ is non-empty, closed and bounded (compactness by Heine-Borel Theorem \citep{borel1895quelques}) and a convex subset of certain Euclidean space.
    \item The best response correspondence $br$ is a set-valued function such that $br$ has a closed graph and $br(\cdot)$ is non-empty and convex. 
\end{enumerate}
Then according to Kakutani's fixed point theorem, a NE exists in an empirical game.

For the first condition, since $\Lambda$ is non-empty, then $\Delta(\Lambda)$ is just the simplex of $\Lambda$ so it is non-empty. 
For the compactness, we need to prove $\Delta(\Lambda)$ is closed and bounded.
First note that $|\Lambda|$ is finite since there are finite number of strategies in the empirical game.
Since $\Delta(\Lambda)$ is the intersection of the closed sets $\mathcal{R}^{|\Lambda|}_+$ and $\{\lambda\in \mathcal{R}^{|\Lambda|}: \sum_{j\in [|\Lambda|]} \lambda_j = 1\}$, $\Delta(\Lambda)$ is closed.
Since $\Delta(\Lambda)$ is a subset of $[0,1]^{|\Lambda|}$, it is bounded.
Then $\Delta(\Lambda)$ is compact.
For the convexity, consider any strategies $s$ and $s'$, and coefficient $\lambda\in(0,1)$, according to the definition of a simplex, $\lambda s + (1-\lambda) \in \Delta(\Lambda)$. 
So $\Delta(\Lambda)$ is a convex set.
We now complete the verification of the first condition.

For the second condition, given a strategy $\sigma\in \Delta(\Lambda)$, define a best-response correspondence to $\sigma$ as 
\begin{align*}
    br(\sigma) & = \argmax\limits_{s'\in\Lambda} u(s', \mu^\sigma) \\
    & = \argmax\limits_{\sigma'\in\Delta(\Lambda)} u(\sigma', \mu^\sigma)\\
    & = \argmax\limits_{\sigma'\in\Delta(\Lambda)} \sum\limits_{s\in\Lambda} \sigma'(s) u(s, \mu^\sigma)
\end{align*}

Due to the compactness of $\Delta(\Lambda)$ and the continuity assumption of $u$, the best-response correspondence $br(\sigma)$ is non-empty. 
To show it is convex, consider two strategies $s_1, s_2\in br(\sigma))$ associated with coefficients $c_1$ and $c_2$ such that $c_1, c_2 \ge 0$ and $c_1 + c_2 = 1$.
Since all optima share the same utility
\begin{displaymath}
    u(s_1, \mu^\sigma) = u(s_2, \mu^\sigma) 
\end{displaymath}
and the definition of the expected utility
\begin{displaymath}
    u(c_1s_1 + c_2s_2, \mu^\sigma) = c_1u(s_1, \mu^\sigma) + c_2u(s_2, \mu^\sigma)
\end{displaymath}
we have $c_1s_1 + c_2s_2 \in br(\sigma)$ and then $br(\sigma)$ is convex.
Note that $br(\sigma)$ is a set-valued function since there could be multiple strategies maximizing the value function, which constitutes a power set of $\Lambda$.

Next, we claim that $Gr(br) := \{(\sigma, br(\sigma)): \sigma\in \Delta(\Lambda), br(\sigma)\in \Delta(\Lambda)\}$ is a closed graph. 
By the Berge's maximum theorem \citep{berge1997topological} and the continuity assumption, the set-valued function $br$ is upper-hemicontinous.
Since $br(\sigma)$ is closed for all $\sigma\in \Delta(\Lambda)$ and $\Delta(\Lambda)$ is a metrizable space, $Gr(br)$ is a closed graph.
This completes the proof of the second condition.
With condition 1 and 2, according to Kakutani's fixed point theorem, a NE exists in an empirical game.

\end{proof}

\subsection{Proof of Theorem~\ref{thm: DO convergence}}

\begin{proof}[Proof of Theorem~\ref{thm: DO convergence}]
Suppose $\sigma^*$ is an empirical NE associated with a population distribution $\mu^*$ such that 
\begin{displaymath}
    \max_{s\in \Lambda}u(s, \mu^*)-u(\sigma^*, \mu^*) = 0
\end{displaymath}
Suppose there is no beneficial deviation can be found at certain iteration, indicating 
\begin{displaymath}
    \max_{s\in S}u(s, \mu^*)-u(\sigma^*, \mu^*) = 0
\end{displaymath}
Then $\sigma^*$ is a NE of the full game. 
Since the strategy space is finite, $\sigma^*$ is always reachable with the worst case where all strategies are included in the empirical game.

\end{proof}

\section{Experimental Results for Multi-population MFGs}\label{app: pred}

For MP-MFGs, we test the performance of iterative EGTA in a 3-population chasing problem \citep{perolat2021scaling}, which closely relates to the game Hens-Foxes-Snakes, where hens, snakes and foxes are chasing cyclically.
The reward structure of this game is shown in table~\ref{tab:hsf}, denoted as $\mathit{R}$.

\begin{center}
\begin{tabular}{ c|c c c } 
  & Hens & Snakes & Foxes \\ 
 \hline
 Hens & (0, 0) & (-1, 1) & (1, -1) \\ 
 Snakes & (1, -1) & (0, 0) & (-1, 1)\\ 
 Foxes & (-1, 1) & (1, -1) &  (0, 0)\\ 
\end{tabular}\label{tab:hsf}
\end{center}
The immediate reward function of population $i\in [N_p]$ is defined as 
\begin{displaymath}
    r_i(x,a,\mu) = -\log(\mu_i(x)) + \sum_{j\ne i} \mu_j(x) \mathit{R}(i, j)
\end{displaymath}

\begin{figure}[!htpb]
\centering
\includegraphics[width=0.9\columnwidth]{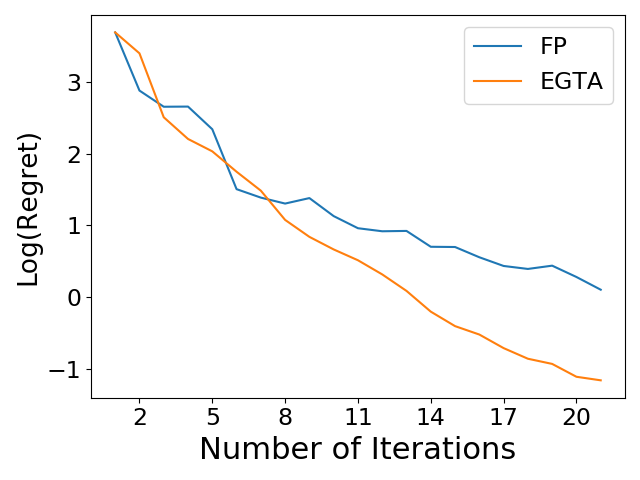}
\caption{Regret curves in multi-population chasing.}\label{fig: pred}
\end{figure}

In Figure~\ref{fig: pred}, we observe that at the early stage of learning, both FP and iterative EGTA can learn the game quickly and improve the stability of strategies.
Iterative EGTA keeps its momentum as learning proceeds while the learning curve of FP becomes flatten over time.
We conjecture that the reason for the two learning curves of EGTA and FP close to each other is the best response target given by FP is not as effective as the one in the single-population setting since FP will not generally converge to NE in MP-MFGs. 

\section{Comparing RD with FP as a Subroutine} \label{app: RD and FP}
\begin{figure}[!htpb]
\centering
\includegraphics[width=0.9\columnwidth]{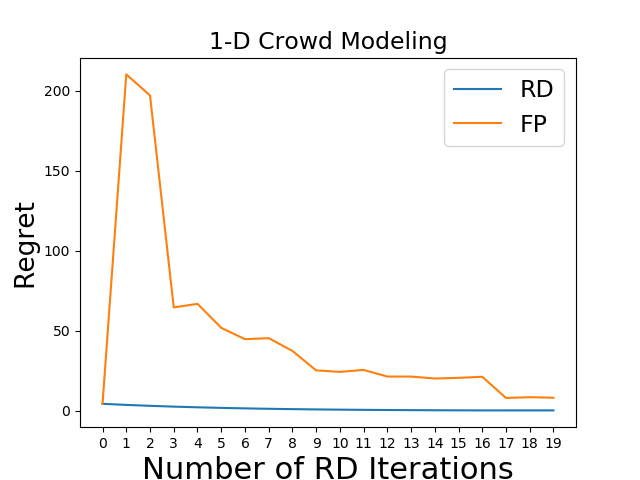}
\caption{RD vs FP.}\label{fig: RD vs FP}
\end{figure}

In our experiments, we observe that both RD and FP can successfully serve as tools for empirical game analysis. 
Although FP has been proved convergence in MFGs by \citet{elie2020convergence} and \citet{perrin2020fictitious}, we observe that RD also empirically converges to NE in our MFGs.
For the subroutine selection, we prefer RD than FP since RD appears to be more stable than FP in some cases (e.g., Fig~\ref{fig:1d RD good} and Fig~\ref{fig:2d RD good}).
We also find that while evaluating each intermediate empirical game, the regret curve of RD is more stable than FP.
Specifically, we consider the empirical game at the final iteration of EGTA, which includes all strategically important strategies.
We apply RD and FP separately to the NE of this empirical game and report the regret curves in Figure~\ref{fig: RD vs FP}.
We observe than given an empirical game, the performance of RD is more stable than that of FP and more importantly RD converges faster than FP.
Therefore, we prefer RD to FP but both are highly effective for strategy generation.

\section{Future Research Directions} \label{app: limitations}
\subsection{Complexity of the Empirical Game Analysis}
In our experiments, we measure the regret with respect to EGTA iterations and show superior performance of iterative EGTA against FP\@. 
However, in terms of running time, we notice that directly applying FP or OMD to the MFGs in OpenSpiel results in faster convergence since in EGTA the analysis of the empirical game (i.e., estimate the payoffs of mixed strategies, update the mixed strategies and compute the corresponding distributions) turns out to be computationally expensive. 
This is because without an explicit payoff matrix, the same strategy or similar strategies and distributions could be evaluated repeatedly.
Besides, computing the induced distribution for every FP update also could be costly. 
In a word, we believe that the acceleration of the empirical game analysis is a crucial step for the application of iterative EGTA for MFGs,
which is a potential research direction in the future. 

In our plots, the reason for measuring the regret with respect to EGTA iterations is based on one assumption in EGTA, that is, it is common that the cost on best response calculation becomes dominant than the empirical game analysis in complex games especially when deep reinforcement learning is deployed.
Based on this assumption, we prefer to build an effective game model with a minimal number of iterations and thus evaluating the performance with respect to the EGTA iterations.
For the same reason, we did not plot the regret curves of OMD \citep{perolat2021scaling} with respect to PSRO iterations since we found that the definition of one iteration in OMD is different from that in PSRO, which could lead to improper comparison.

For MP-MFGs, as we observed in the multi-population chasing experiments, since FP will not generally converge to a NE in the multi-population setting, using FP for the empirical game analysis in EGTA may affect the learning performance.
Therefore, selecting an effective best response target is also one future research direction.

\subsection{Re-evaluating Strategies in Finite Games}
A key motivation for using MFGs is that the MFG model dramatically simplifies game learning compared to directly solving the corresponding finite game.
Meanwhile, the solution of an MFG approximates the NE of the finite game.
Since the approximation induces some errors, one future research direction is how to reduce the error while applying the MFG solution to the finite game.
It is apparent that the error depends on many factors (e.g., the number of players and the game size of a finite game).

A key distinction between EGTA and other learning dynamics is that the constructed empirical game model incorporates a set of strategies, which are considered strategically important to understand the game.
The construction of the set of strategies is called the \textit{strategy exploration} problem in EGTA, which aims to construct effective models with minimal iteration.
The iterative EGTA can be viewed as an approach for strategy exploration in MFGs.

Based on this feature of EGTA, to reduce the approximation error of the MFG solution in the corresponding finite game, one potential approach is to conduct strategy exploration in the MFG while re-evaluating the generated strategies in the finite game.
This takes the virtue of the MFG model for fast strategy exploration (i.e., quickly obtain strategically important strategies of a game) as well as improving the accuracy of the empirical game model for the finite game.
This approach is also compatible with the factors that affect the approximation errors.
For example, an MFG solution could behave much worse in a ten-player symmetric game than in a one-hundred-player symmetric game. 
In this case, re-evaluating the empirical game model in the ten-player game could improve the accuracy of the empirical game model and provide stable solutions.
Note that the re-evaluation does not necessarily require evaluating all profiles in the empirical game by utilizing the symmetry and certain techniques \citep{gemp2021sample}.
Due to the potential benefits of this approach for solving a finite game, we propose this as a future research direction.

\end{document}